\documentclass[twocolumn,english,aps,prl]{revtex4-1}
\usepackage[T1]{fontenc}
\usepackage[latin9]{inputenc}
\usepackage{graphicx}

\makeatletter
\usepackage{babel}

\usepackage{babel}

\makeatother

\usepackage{babel}
\begin{document}

\title{
Magnetism of gadolinium: a first-principles perspective}

\author{L. Oroszl\'any$^{1}$, A. De\'ak$^{1}$, S. Khmelevskyi$^{1}$, L. Szunyogh$^{1,2}$}

\affiliation{$^{1}$ Department of Theoretical Physics, Budapest University of
Technology and Economics, Budafoki \'ut 8, H-1111 Budapest, Hungary}

\affiliation{$^{2}$ MTA-BME Condensed Matter Research Group, Budapest University
of Technology and Economics, Budafoki \'ut 8, H-1111 Budapest, Hungary}
\begin{abstract}
By calculating the spectral density of states in the ferromagnetic
(FM) ground state and in the high temperature paramagnetic (PM) phase
we provide the first concise study of finite temperature effects on
the electronic structure of the bulk and the surface of gadolinium
metal. The variation of calculated spectral properties of the Fermi
surface and the density of states in the bulk and at the surface are
in good agreement with recent photoemission experiments performed
in both ferromagnetic and paramagnetic phases. In the paramagnetic
state we find vanishing spin splitting of the conduction band, but
finite local spin moments both in bulk and at the surface. We clearly
demonstrate that the formation of these local spin moments in the
conduction band is due to the asymmetry of the density of states in
the two spin channels, suggesting a complex, non-Stoner behavior.
We, therefore, suggest that the vanishing or nearly vanishing spin
splitting of spectral features cannot be used as an indicator for
Stoner-like magnetism. 
\end{abstract}
\maketitle
Pure hcp Gd metal is perhaps the most investigated strongly correlated
metallic system, where the well localized magnetic moment of the $4f$-shell
interacts with the itinerant electrons of the conduction band.\cite{Jensen,LinbaumRotter}
The half-filled $4f$-shell of Gd possesses a spin-moment of $7\mu_{B}$,
and it is energetically well separated from the conduction band of
the $spd$-electrons. In the FM ground state the ordered local moments
induce a spin splitting of about $1\,\mbox{eV}$ of the conduction
band resulting in an additional spin magnetic moment of $0.6\,\mu_{B}$
\cite{Roeland-exp-magmom}.

Modern electronic structure theory describes adequately the ground
state of Gd, if the strong correlation in the $4f$-shell is treated
in some, even simple, manner. In pioneering works it has been shown
that the LSDA+U method provides a good description of the Gd ground
state in bulk \cite{Anisimov-1-LDAU,Shick-1-LDAU} and on the (0001)
surface \cite{Shick-2-LDAU-surf}. Later studies\cite{Blugel-Surf,Hafner-Surf}
have confirmed this conclusion. Moreover, recent angle-resolved photoemission
(ARPES) measurements of the Fermi surface in the FM phase of the bulk
\cite{Dobrich-1-ARPES} have been successfully reproduced using the
LSDA+U methodology\cite{Abdelouahed-1-GGAU-electronic}. Despite more
sophisticated treatments of the correlated $4f$-shell in Gd \cite{Temmerman-Nature-SIC2,Mirhosseini-SIC,Kotani-GW},
the considerable splitting of the Gd conduction band could also be
obtained in a remarkably simple spin-polarized open-core approach.\cite{Sandratskii-1}

Owing to the strong localization of $4f$-states, the finite temperature
magnetism in hcp Gd seems to be well described by the Heisenberg model,
and it may be regarded as a model Heisenberg system among metallic
ferromagnets. The magnetic critical temperature of hcp Gd can indeed
be estimated reasonably well in terms of a Heisenberg model and calculating
the exchange constants using either the open-core \cite{Turek-Kudorovski,SK-Gd-magnetic-JPCM,Sandratskii-Kubler},
the LSDA+U \cite{Antropov-Jij} or the self-interaction correction
(SIC) approach \cite{Mirhosseini-SIC}.

However, despite the successful understanding of the ground state
and finite temperature magnetism of Gd, the fundamental issue concerning
the interaction between localized $4f$-moments and itinerant electrons
has been the subject of a heated debate.\cite{Dowben19971,Nolting-DLM,Maiti-1-SPES,SK-Gd-non-Stoner,Sandratskii-1}
Earlier photoemission (PES) experiments\cite{Bongsoo-PES-M8} have
shown that the spin splitting of the conduction band vanishes near
the Curie temperature. This gives rise to a simple interpretation
that the $spd$-bands experience a spin polarization due to the (average)
magnetization of the localized $4f$-electrons, a hallmark of Stoner
magnetism \cite{PhysRevB.45.7272-MAITI7,Bongsoo-PES-M8,PhysRevLett.77.5138-MAITI9,PhysRevLett.83.3017-MAITI10,PhysRevLett.84.5624-MAITI11}.
However, more precise analyses of later PES measurements\cite{Maiti-2-SPES,Dowben-SPES,Fedorov-1-SPES}
seemed to provide clear evidence that the spin splitting of majority
and minority spin channels remains finite in the PM state, although
very small, being on the verge of the PES resolution. From this observation
it was concluded that the Stoner model cannot describe the magnetism
of bulk Gd.\cite{Maiti-1-SPES}

Contradictory theoretical interpretations were also given based on
\emph{ab initio} determined model parameters for the description of
the interaction between the localized $4f$-moments and the conduction
band. For example the authors of Ref.\ \cite{Nolting-spectralfunction}
derived a spin-mixing behavior of the conduction band in agreement
with the experimental conclusions by Maiti \emph{et al.}\ \cite{Maiti-1-SPES,Maiti-2-SPES},
whereas later on the same authors proposed a theory predicting Stoner
behavior \cite{Nolting-DLM} in line with old measurements by Kim
\emph{et al.}\ \cite{Bongsoo-PES-M8}. Fully \emph{ab initio} attempts
to resolve the above theoretical contradiction used the disordered
local moment approximation (DLM) to model finite temperature magnetic
disorder \cite{SK-Gd-non-Stoner} or employed a study of non-collinear
spin-configurations.\cite{Sandratskii-1} Although both of these studies
were based on the open-core treatment of the $4f$-shell, they indicated
a spin-mixing behavior of the conduction band. The DLM calculation
\cite{SK-Gd-non-Stoner} predicted a small but finite splitting of
the majority and minority spin channels and a corresponding finite
value of the local moment due to the $spd$-band in the paramagnetic
state. The non-collinear calculation \cite{Sandratskii-1} clearly
showed that the conduction band moment is strongly coupled to the
direction of the local $4f$-moment and cannot be treated as an independent
magnetic degree of freedom.

In recent years the study of ultrafast magnetic processes induced
by femtosecond laser pulses has been fueling a reinvigorated discussion
concerning the nature of the interaction between $4f$-states and
the conduction band moments in pure Gd and its compounds. \cite{Koopmans2010,Carley-1-laserexcit,Wietstruk}
A crucial point in the simulations of these processes lies in the
assumption of treating $5d$- and $4f$-moments as independent magnetic
degrees of freedom.\cite{Radu2011-GdFe,NowakOppeneer-df-coupling}
An elaborated discussion of this topic was recently given by Sandratskii.\cite{Sandratskii-1}

Recent ARPES \cite{Dobrich-1-ARPES,Dobrich-2-ARPES} and positron
annihilation spectroscopy \cite{Dugdale-positron-annihilation,Fretwell-positron-annihilation,Crowe-positron-annihilation}
experiments were able to observe differences of the Fermi surface
topology between the FM and PM state. However, the current theoretical
understanding of the PM electronic structure of the Gd conduction
bands is far from being satisfactory. The Fermi surface of pure bulk
Gd in the PM state has been calculated for several lattice constants
using self-interaction correction to the LSDA \cite{Temmerman-Nature-SIC2}.
The main features of the derived PM Fermi surface seem to be consistent
with the later measured one \cite{Dobrich-2-ARPES}. However, so far
there has been no consistent theoretical description of the Fermi
surfaces in the FM and PM state and the evolution of spectroscopic
properties from bulk to surface within the same framework.

An equivalently important issue is that the main experimental methods
that provide us with precise information concerning the electronic
structure of the Gd conduction band are surface sensitive. The presence
of the surface leads to essential changes in the electronic structure
of the terminating layers compared to the bulk. Moreover, the spin
splitting of the surface states has been regarded as an important
source of the information concerning the magnetic behavior of the
conduction electrons. \cite{Dowben19971} It is therefore demanding
to disentangle the signatures of the surface state from those of the
bulk.

In this Letter we address both issues, namely the changes of the electronic
structure of hcp Gd induced by finite temperature magnetic disorder
and the variation of the Fermi surface near the Gd (0001) surface
termination. We employ the well established LSDA+U methodology for
the treatment of the half filled $4f$-shell and calculate bulk and
layer resolved Fermi surfaces, as well as densities of states in the
FM and the DLM state. We show that the high temperature behavior of
the conduction band is governed by spin-mixing both in the bulk and
near the surface. The comparison of our results with PES \cite{Maiti-1-SPES,Maiti-2-SPES}
and ARPES studies \cite{Dobrich-1-ARPES,Dobrich-2-ARPES} suggests
a good agreement for both FM and PM phases with regard to bulk and
surface properties. We also find that electronic states associated
with a presence of the surface decay faster towards the bulk in the
paramagnetic phase than in the ferromagnetic state.

We performed self-consistent calculations by using the fully relativistic
screened Korringa--Kohn--Rostoker (KKR) Green's function method with
the atomic sphere approximation, that allows the study of layered
systems and surfaces.\cite{KKRBOOK} The strong correlation of the
localized $4f$-states was treated within the framework of the LSDA+U
approach\cite{Anisimov-1-LDAU,Shick-1-LDAU} as implemented within
the KKR method.\cite{Ebert-SPRKKR-LDAU} The paramagnetic phase is
treated in terms of the relativistic disordered local moment (R-DLM)
method,\cite{Staunton-RDLM,Deak-RDLM} in which the coherent potential
approximation (CPA) is employed to model disorder in local spin orientations.
\cite{Gyorffy-DLM} 
The application of these methods allows us to treat the correlations
in the $4f$-shell on equal footing in FM ground state and finite
temperature PM phase. The spectral density of states (SDOS) in the
FM phase is directly related to the KKR Green's function, while in
the PM phase the SDOS is evaluated from the configurationally averaged
Green's function\cite{Weinberger-1-CPA-spectralfunct,KKRBOOK}. For
all the calculations we considered a hexagonal closed packed lattice
with the experimental value of the $c/a$ ratio of $1.5904$, while
the lattice constant was optimized to $a=3.450\mbox{ \AA}$. The optimization
was performed with the commonly used $U=6.7\mbox{ eV}$ and $J=0.7\mbox{ eV}$
values, \cite{Anisimov-1-LDAU} and the LSDA parametrization from
Ref.\ \cite{PZ-LSDA}.

\begin{figure}[ht]
\includegraphics[width=1\columnwidth]{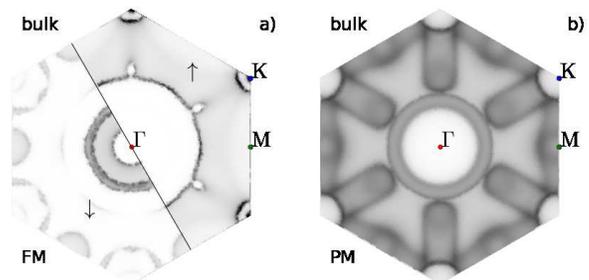}

\protect\caption{a) Spin resolved spectral DOS of the ferromagnetic phase and b) total
spectral DOS of the paramagnetic phase of bulk gadolinium at the Fermi
energy as projected to the 2D Brillouin zone. Darker colors represent
larger values of the spectral DOS. \label{fig:bulk-sAk-FM}}
\end{figure}

\begin{figure}[t]
\includegraphics[bb=60bp 50bp 1050bp 800bp,clip,width=1\columnwidth]{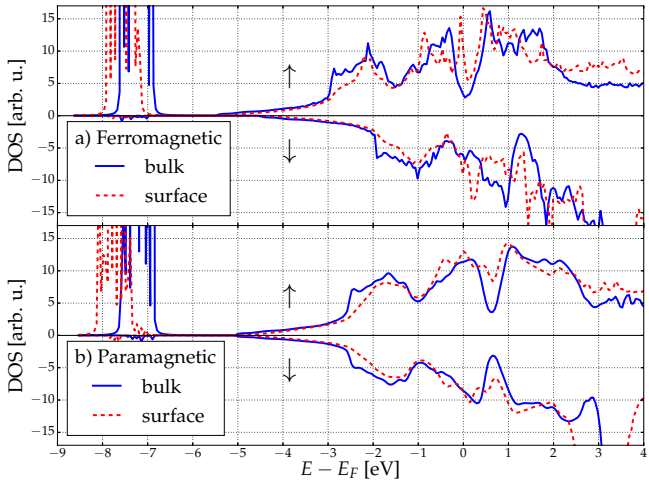}

\caption{Spin resolved densities of states in Gd bulk and in the top layer
of Gd(0001) surface in the a) ferromagnetic and b) paramagnetic phase.
\label{fig:Dos}}
\end{figure}

The spin resolved SDOS of the FM bulk and the total SDOS of the PM
bulk is depicted in Fig.\ \ref{fig:bulk-sAk-FM}, projected onto
the hexagonal two-dimensional (2D) Brillouin zone. 
As can be clearly seen in Fig.\ \ref{fig:bulk-sAk-FM} a), the Fermi
surface in the case of the FM phase shows a significant spin splitting
in momentum space with concentric features around both the $\Gamma$
and $K$ points of the Brillouin zone, in agreement with ARPES experiments.\cite{Dobrich-2-ARPES}
The minority ($\downarrow$) components form a tight cylinder around
the $\Gamma$ point surrounded by a second cylinder consisting entirely
of majority ($\uparrow$) spin-states. A similar, but much narrower
feature can be observed around the $K$ points with the order of the
spin channels reversed, that is the majority states form the inner
and the minority components the outer cylinder. In the paramagnetic
phase the splitting disappears, only a single cylindrical feature
is present around the $\Gamma$ and $K$ points, nicely recovering
the experimental observations.\cite{Dobrich-2-ARPES} These experimental
and theoretical findings seem to lead to the simple conclusion that
a vanishing spin splitting in the paramagnetic phase is an indication
of Stoner behavior, 
where the spin splitting depends on the total magnetization of Gd.

The densities of states for the bulk and surface in the FM and PM
phases are depicted in Fig.~\ref{fig:Dos}. In the FM phase, similarly
to the results of previous works \cite{Anisimov-1-LDAU,Shick-1-LDAU,Shick-2-LDAU-surf,Abdelouahed-1-GGAU-electronic},
the $4f$-electrons move away from the Fermi level with an exchange
splitting of $\sim11\,\mbox{eV}$. Note that the relatively large
dispersion of about $0.8\,\mbox{eV}$ of the majority $4f$-states
is due to spin-orbit splitting of these states. Clearly, in the PM
phase very similar features of the density of $4f$-states can be
found. 
The conduction electrons, dominated by $5d$-states, are characterized
by a spin splitting on the order of $1\,\mbox{eV}$ in the FM state.
The magnitude of this splitting can be inferred from Fig.~\ref{fig:Dos}a),
for example by comparing the onset of the majority $d$ electrons
at roughly $-3\,\mbox{eV}$ below the Fermi energy to the onset of
minority $d$-electrons at $-2\,\mbox{eV}$. Another measure of the
spin splitting would be to consider the shift of the minimum in the
majority-spin DOS at the Fermi energy to roughly $1.2\,\mbox{eV}$
above the Fermi level in the minority-spin DOS. The spin splitting
of bulk states in the PM phase is somewhat harder to quantify with
the above measures. A comparison of the position of the peak near
the Fermi energy for the two spin channels suggests a largely reduced
splitting of the order of $0.1\,\mbox{eV}$, as can be assessed from
Fig.~\ref{fig:Dos}~b). At this point one might again conclude that
an almost vanishing spin splitting in the PM phase indicates Stoner
behavior.

In the FM phase we calculated a total spin moment of $7.77\,\mu_{B}$,
that is, the polarization of the $spd$-band amounts to $0.77\,\mu_{B}$.
In the PM phase the total spin magnetic moment is reduced to $7.41\,\mu_{B}$,
which means that a considerable polarization of the conduction band
still persists. These values are in agreement with recent disordered
local moment calculations\mbox{\cite{SK-Gd-non-Stoner,Temmerman-Nature-SIC2}}
and also with studies performed for non-collinear magnetic configurations
\mbox{\cite{Sandratskii-1}}, where the $4f$-electrons are treated
as part of the core. The finite value of the local moment related
to the conduction band in the PM phase is indeed at the heart of the
controversy, since there is (nearly) no splitting in the spin channels
either in momentum space or in the energy resolved spectra. 

\begin{figure*}[ht]
\includegraphics[width=1\textwidth]{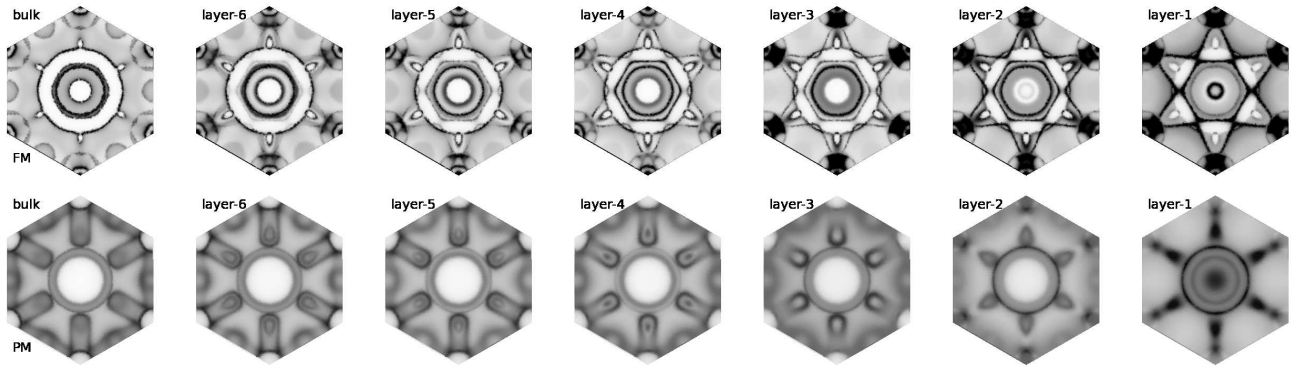} \protect\caption{Spectral densities of states evaluated at the Fermi energy in the
ferromagnetic (top row) and paramagnetic (bottom row) phases for Gd
bulk and the first six layers below a Gd(0001) surface. \label{fig:surf-dos}}
\end{figure*}

The density of states at the surface, shown in Fig.~\ref{fig:Dos},
has two main characteristics. First, in accordance with previous results
\cite{Shick-2-LDAU-surf} and experimental observation,\cite{Dowben-experiment-shift-4f-surface}
the localized $4f$-states experience a down-shift in energy. This
feature is slightly more pronounced in the PM phase than in the FM
phase. The second feature, again in line with experiments, \cite{Maiti-1-SPES,Maiti-2-SPES,Fedorov-1-SPES}
is the appearance of new states absent in the bulk. The surface states
in the FM phase near the Fermi energy are highly of majority-spin
character. In the PM phase, we still detect a track of such surface
states near the Fermi energy, but with a reduced spin polarization
as observed in experiments.\cite{Maiti-1-SPES,Maiti-2-SPES} The spin
magnetic moment per atom on the surface in the FM phase is found to
be slightly reduced, compared to the bulk value, to $7.74\,\mu_{B}$.
On the other hand, in the PM phase the magnetization slightly increases
to $7.48\,\mu_{B}$ on the surface. Therefore, similarly to the case
of bulk, the vanishing splitting of the spectral density of the surface
states in the PM phase can not be regarded as a herald of Stoner magnetism.
It is the asymmetry of the spectral weight for the two spin channels
that is responsible for the finite residual local moment of the conduction
band in the PM phase, suggesting a spin-mixing non-Stoner behavior.

As we stated before, most spectroscopic methods are surface sensitive.
So far it is not clear how deep the effects of surface termination
on the electronic spectra extend in to the bulk. To elucidate an answer
for this question, in Fig.~\ref{fig:surf-dos} we present the layer
resolved spectral DOS at the Fermi energy for the first six layers
below the (0001) surface and compare them with the bulk spectral function.
The most pronounced effects of the surface can be seen on the first
(surface) layer. In both phases a high intensity feature is present
in the vicinity of the $\Gamma$ point which is absent in the subsequent
layers. In the FM phase an enhancement of the spectral density around
the $K$ points and along the lines connecting equivalent $K$ points,
forming a hexagram like structure, can also be attributed to the surface
states. These highly spin polarized features gradually fade in deeper
layers. In the PM phase no such distinctive spectral features are
present near the surface. We merely observe a slight modulation in
the SDOS along the $\Gamma-K$ direction in the region out of the
disk of zero weight characterizing the bulk. Our calculations show
that the penetration depth of the surface states is different in the
two phases. A bulk like spectral density is nearly recovered in the
3rd layer from the surface in the PM phase, while in the FM phase
the SDOS even at the 5th-6th layers below the surface contain surface
associated features. Thus one can deduce that the surface effects
are more pronounced in the FM phase than in the high temperature PM
state.

In conclusion, we provided a coherent first-principles study of the
electronic and magnetic properties of the bulk and the surface of
Gd metal, both in the ferromagnetic ground state and in the high temperature
paramagnetic phase. Our results reveal the root of the existing controversy
concerning the nature of magnetism related to the conduction band
of Gd. A finite, though reduced, value of the spin moment related
to the conduction band in the paramagnetic phase both in the bulk
and for the surface implies a non-Stoner magnetism for these electrons.
We argue that the asymmetry of the density of states with respect
to different spin channels can be regarded as an indicator for this
behavior rather than the spin splitting of the bulk conduction band
or of the surface states, as it has been judged in many previous investigations.
The asymmetry of the conduction spin-bands is generated by stable
local $4f$-moments irrespective of their directions and it is the
consequence of the formation of a common band, i.e.~of a spin-mixing
state.

\acknowledgments This work was supported by the European Union under
FP7 Contract NMP3-SL-2012-281043 FEMTOSPIN, and the Hungarian Scientific
Research Fund under grant No.~K108676 and K84078.

 \bibliographystyle{apsrev}

\end{document}